\documentclass[aps,nofootinbib]{revtex4}

\usepackage{epsfig}
\usepackage{epsf}
\usepackage{bm}                 
\usepackage{amsmath}
\usepackage{amssymb}
\usepackage{dcolumn}
\usepackage{graphicx}
\usepackage{xcolor}
\usepackage[small,bf, format=plain, width=.75\textwidth, justification=RaggedRight]{caption}[2008/04/01]
\input amssym.def
\input amssym

\font\bbold=msbm10

\newcommand{\BR}{\mbox{\bbold R}}
\newcommand{\BC}{\mbox{\bbold C}}

\begin{document}

\title{Fundamental Limits on the Speed of Evolution of Quantum States}

\author{Ulvi Yurtsever} \email{ulvi@phys.lsu.edu}
\affiliation{MathSense Analytics, 1273 Sunny Oaks Circle, Altadena, CA 91001 and
\\ Hearne Institute for Theoretical Physics, Louisiana State University,
Baton Rouge, LA 70803}

\date{\today}

\begin{abstract}
This paper reports on some new inequalities of
Margolus-Levitin-Mandelstam-Tamm-type involving the speed of
quantum evolution between two orthogonal pure states. The
clear determinant of the qualitative behavior of
this time scale is the statistics of the
energy spectrum. An often-overlooked
correspondence between the real-time behavior of a quantum system and the
statistical mechanics of a transformed (imaginary-time) thermodynamic
system appears promising as a source of qualitative
insights into the quantum dynamics.
\end{abstract}


\maketitle
\thispagestyle{plain}

\noindent {\bf Motivation:}

As quantum information processors evolve in architectural
complexity, it will be increasingly important to develop qualitative,
architecture-independent methods that can answer fundamental questions
about processor properties as a function of growing complexity without recourse
to detailed, large-scale simulations,
and independently of the details of processor architecture. The key questions
will be speed of quantum evolution (analogous to clock speed),
decoherence rates, and stability through control of decoherence.

Basic limits on the speed of quantum evolution
can be deduced directly from the Schr\"odinger equation. Consider,
for example, a system with Hamiltonian $H$ in a pure
quantum state evolving in time according to
\begin{equation}
| \psi_t \rangle = e^{-i H t} | \psi_0 \rangle \; ,
\end{equation}
where we rescaled the time parameter $t$ in inverse-energy units
as $t \equiv \rm{time}/\hbar$ (or, equivalently, set $\hbar = 1$).
Expanding in the energy eigenbasis $|E_n \rangle$ of the Hamiltonian we obtain
\begin{equation}
\Phi(t) \equiv \langle \psi_0 | \psi_t \rangle
= e^{-i E_0 t} \sum_{n=0}^{\infty} |c_n |^2 e^{-i (E_n - E_0 ) t}
 \; ,
\end{equation}
where $| E_0 \rangle$ is the lowest-energy (ground) state of $H$
(so $E_n - E_0 \geqslant 0 \; \forall n$). According to Eq.\,(2)
\begin{equation}
\mbox{Re} \, [e^{i E_0 t} \Phi (t) ]
=\sum_{n=0}^{\infty} |c_n |^2 \cos ((E_n - E_0 ) t ) \; ,
\end{equation}
and making use of the elementary inequality (see Fig.\,1)
\begin{equation}
\cos x \geqslant 1 -\frac{2}{\pi}(x + \sin x ) \; \; \; \;  \forall \; x \geqslant 0 \; ,
\end{equation}
\begin{eqnarray}
\mbox{Re} \, [ e^{i E_0 t} \Phi (t) ]
& \geqslant & \sum_{n=0}^{\infty} |c_n |^2 
\left( 1- \frac{2}{\pi} (E_n -E_0 ) t - \frac{2}{\pi} \sin((E_n - E_0 )t) \right) \nonumber \\
& = & 1- \frac{2 t}{\pi} \langle H- E_0 \rangle + \frac{2}{\pi} \, \mbox{Im} \,
[ e^{i E_0 t} \Phi (t) ] \; ,
\end{eqnarray}
where $\langle F \rangle \equiv \sum_{n=0}^{\infty} |c_n |^2 F_n
= \langle \psi_0 | F | \psi_0 \rangle $ denotes the expectation value
of an observable $F$ in the initial state $| \psi_0 \rangle$.
If $T_0$ denotes the first zero of the overlap $\Phi(t)$
(i.e.\ the earliest time $t$ at which $\psi_0$ evolves to an orthogonal state),
after restoring to natural time units
Eq.\,(5) yields the inequality
\begin{equation}
T_0 \geqslant \frac{\pi \hbar}{2 \langle H-E_0 \rangle} \; .
\end{equation}
\begin{figure}[htp]
\vspace{0.0in}
\hspace{3.0in}
\begin{center}
\epsfig{file=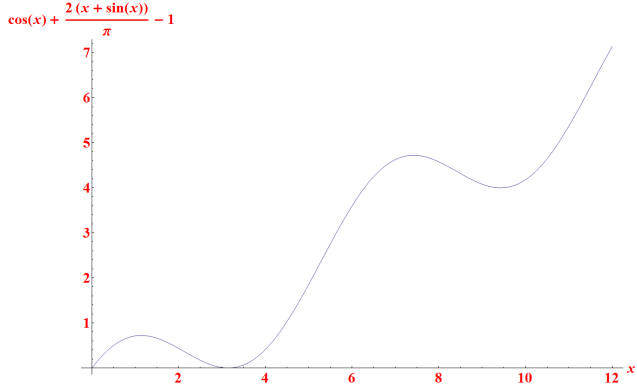,height=2.1875in,width=3.4027in,angle=0}
\end{center}
\vspace{0.0in}
\caption[figure]{\label{fig:figure1}
Proof of the inequality $\cos(x) \geqslant 1 - (2/\pi ) (x+\sin (x))$.}
\vspace{0.0in}
\end{figure}
\noindent On the other hand, the Schr\"odinger equation in the Heisenberg form for an operator $A$
\begin{equation}
\frac{d A}{dt} = \frac{1}{i \hbar} [H,A]
\end{equation}
combined with the uncertainty inequality (for any state vector $| s \rangle$)
\begin{equation}
(\Delta A)^2 (\Delta H )^2 \geqslant \frac{1}{4} | \langle s | [A,H] | s \rangle |^2
\end{equation}
where $(\Delta A)^2 \equiv \langle (A - \langle A \rangle )^2 \rangle 
= \langle s | A^2 |\ s \rangle - \langle s | A | s \rangle^2$, gives rise to the
Mandelstam-Tamm-Margolus-Levitin inequality~\cite{mantamm}
\begin{equation}
T_0 \geqslant \frac{\pi}{2} \frac{\hbar}{ \Delta H} \; .
\end{equation}

Lower bounds on the first zero $T_0$ probe the ``long-time" behavior of the overlap
$\Phi (t) =  \langle \psi_0 | \psi_t \rangle$, and
can be interpreted as ``speed limits" on the rate of quantum
evolution~\cite{hookstuff}. On the
other hand, the short-time behavior
of $\Phi (t)$ is completely determined by the variance $\Delta H$ since it can be shown
easily that
\begin{equation}
|\Phi (t)|^2 = 1 - \frac{(\Delta H)^2}{\hbar^2} t^2 + O(t^3 ) \;.
\end{equation}
As a quick application, Eq.\,(10) implies that the quantum Zeno effect~\cite{zeno} for
the initial state $|\psi_0 \rangle$,
which requires the condition $\lim_{n \rightarrow \infty} |\Phi (t/n)|^{2 n} = 1$,
is in principle always present as long as the variance $\Delta H$ is finite. More
precisely, if $n$ projective measurements of the operator
$| \psi _0 \rangle \langle \psi _0 |$ are performed successively
at equal intervals $t/n$, the enhanced probability of finding the system in the initial
state  $|\psi_0 \rangle$ at time $t$ is given by
\begin{equation}
|\Phi (t/n)|^{2 n}  \approx e^{-\frac{(\Delta H)^2}{\hbar^2} \frac{t^2}{n}} \;.
\end{equation}
Consequently, this ``stasis" probability can be made close to unity provided
\begin{equation}
n  \, \gtrsim \, \frac{(\Delta H)^2}{\hbar^2} t^2 
\, \gtrsim \, \left( \frac{t}{T_0 } \right)^2 \; .
\end{equation}

While the Zeno effect is significant for applications involving quantum control
of decoherence, long-time behavior of the overlap $\Phi (t)$ has significance for
exploring ultimate limits
on the speed of quantum information processing. In fact, limits on the speed of quantum
evolution are connected to another widely known fundamental
limit on the power of information processing, the holographic entropy bound, as suggested
by the following arguments: Consider a general
physical system of approximate spatial size $L$ (hence
surface area $L^2$) and total energy $E$. By the discussion above, there is
a fundamental bound given by an inequality of the type Eq.\,(6) or Eq.\,(9)
on the time $\tau$ during which a quantum state of the system
can evolve to an orthogonal state:
\begin{equation}
\tau > \frac{\hbar}{E} \; .
\end{equation}
Let $S$ be the entropy the system, which is,
equivalently, the Boltzmann constant $k_B$ times the number of
(classical) bits of information that can be stored. Since a physical
signal propagating across the system can be used to interact with
each classical ``bit" successively and change its state, the shortest time in which this
could be done, $(S/k_B ) \tau$, must be no longer than the light crossing time
across the system:
\begin{equation}
\frac{S}{k_B} \tau < \frac{L}{c} \; .
\end{equation}
Combining Eq.\,(14) with Eq.\,(13) yields
\begin{equation}
\frac{S}{k_B} < \frac{L}{c \tau} < \frac{E L}{\hbar c} \; .
\end{equation}
The entropy bound Eq.\,(15) (which was first discovered by
Bekenstein~\cite{jacobb}) can also be derived, independently, by an argument based on
the system's formation history. The speed-of-evolution bound Eq.\,(13) imposes an
ultimate limit on the time rate of entropy change:
\begin{equation}
\frac{1}{k_B} \left | \frac{\partial S}{\partial t} \right | < \frac{1}{\tau} < \frac{E}{\hbar} \; .
\end{equation}
But no matter what the system's actual formation history and its final state are,
a state of zero entropy must be reachable from that final state within a time
on the order of the light crossing time $L/c$. Hence
the Bekenstein bound Eq.\,(15) on the final entropy follows once again, this time from Eq.\,(16).
Since black-hole formation (gravitational collapse)
imposes the Schwarzschild limit $E< c^4 L/(2\, G)$ on the
maximum energy $E$ of the system, Eq.\,(15) implies the holographic entropy bound~\cite{heb}
\begin{equation}
\frac{S}{k_B} < \frac{c^3 L^2}{2 \, \hbar \, G} = \frac{1}{2}
\frac{L^2}{{l_p}^2} \; ,
\end{equation}
where $l_p \equiv \sqrt{\hbar G / c^3}$ is the Planck length.

~

{\noindent \bf New inequalities:}

A key observation about the overlap function
$\Phi (t)$ (Eq.\,(2)) is that it can be expressed as the Fourier
transform (or ``characteristic function") of a positive probability
distribution function $\rho (E)$, characterizing the
energy spectrum of the system in the initial quantum state $| \psi_0 \rangle$:
\begin{equation}
\Phi (t) = \int e^{-i t E} \rho(E) \, dE \; ,
\end{equation}
where again we used the rescaled time parameter
$t \equiv \rm{time}/\hbar$.
For example, if the Hamiltonian $H$ has a discrete spectrum
$\{ E_i \}$, the state $| \psi_0 \rangle$ can be expanded as
\begin{equation}
|\psi_0 \rangle = \sum_j c_j | E_j \rangle \; ,
\end{equation}
and the distribution function  and the overlap have the discrete forms
\begin{equation}
\rho (E) = \sum_j |c_j |^2 \delta (E - E_j ) \; , \; \; \; \; \; \;
\Phi (t) = \sum_j |c_j |^2 e^{-i E_j t} \; .
\end{equation}
More generally, using the spectral theorem~\cite{specthm},
the Hamiltonian $H$ can be expressed in the form
\begin{equation}
H=\int E \, dP(E) \; ,
\end{equation}
where $dP(E)$ denotes integration with respect to a projection-valued
measure on the Hilbert space, and
the energy probability distribution (``density of states") $\rho (E)$
can be defined via the identity
\begin{equation}
\rho(E)\, dE = \langle \psi_0 | \, d P(E) \, | \psi_0 \rangle \; .
\end{equation}
From a practical point of view, the distribution function
$\rho (E)$ is a key design parameter since it is relatively
easy to manipulate. This distribution is related to the overlap function 
$\Phi (t)$ via the inverse Fourier transform
\begin{equation}
\rho (E) = \frac{1}{2 \pi} \int_{-\infty}^{\infty} e^{i E t} \Phi (t) \, dt \; .
\end{equation}

It is useful to extend the function $\Phi (t)$ to the complex domain;
in fact, we will assume a Paley-Wiener~\cite{katznelson}
condition on the distribution $\rho (E)$
such that $\Phi (t)$ as defined by Eq.\,(18) is an entire analytic function on the complex
plane $t \in \BC$. For example, this is the case if $\rho (E)$ is of compact support,
or, more generally, falls off faster than any exponential as $E \rightarrow \pm \infty$.
In either case, one can reasonably expect such Paley-Wiener conditions
to hold for most physical systems.

Denoting complex time with the commonly-used symbol $z$,
Eq.\,(18) implies that the analytic function $\Phi (z)$ has the power series
expansion
\begin{equation}
\Phi (z) = 1+ \sum_{n=1}^{\infty} \frac{(-i)^n \langle E^n \rangle}{n!} z^n \; ,
\end{equation}
which is a rephrasing of the generating-function identities
\begin{equation}
\langle E^n \rangle = i^n \left. \frac{\partial^n \Phi}{\partial t^n}\right|_{t=0}
= i^n \, \Phi^{(n)}(0) \; .
\end{equation}
Consequently, the function $\log \Phi (z)$ can be expanded in a power series
around $z=0$:
\begin{equation}
\log \Phi (z) = \sum_{n=1}^{\infty} \gamma_n  z^n \; ,
\end{equation}
where
\begin{eqnarray}
\gamma_n & \equiv &
\sum_{k=1}^{n} \; \frac{(-1)^{k-1}}{k}
   \sum_{\substack{l_i \geqslant 1 \\ l_1+l_2+\cdots +l_k=n}}^n
   \frac{(-i)^{l_1} \langle E^{l_1}\rangle}{l_1 !} \; \frac{(-i)^{l_2}\langle E^{l_2}\rangle}{l_2 !} \;
   \cdots \; \frac{(-i)^{l_k}\langle E^{l_k}\rangle}{l_k !} \nonumber \\
& = &
    \frac{(-i)^n}{n !} \;
    \sum_{k=1}^{n} \; \frac{(-1)^{k-1}}{k} \sum_{\substack{l_i \geqslant 1 \\ l_1+l_2+\cdots +l_k=n}}^n
    \binom{n}{l_1 \; l_2 \; \cdots \; l_k} \; \langle E^{l_1} \rangle \; \langle E^{l_2} \rangle \cdots
    \langle E^{l_k} \rangle \; .
\end{eqnarray}
Clearly, $T_0$ cannot be less than the radius of convergence of the power
series Eq.\,(26), which gives us our first new inequality:
\begin{equation}
T_0 \geqslant \frac{\hbar}{\lim_{n \rightarrow \infty} | \gamma_n |^{\frac{1}{n}}} \; ,
\end{equation}
where
\begin{equation}
\gamma_n = (-i)^n \; \sum_{k=1}^{n} \; \frac{(-1)^{k-1}}{k}
   \sum_{\substack{l_i \geqslant 1 \\ l_1+l_2+\cdots +l_k=n}}^n
   \frac{ \langle E^{l_1}\rangle \; \langle E^{l_2}\rangle  \cdots \; \langle E^{l_k}\rangle}
   {l_1 ! \; l_2 ! \; \cdots \; l_k !} \; .
\end{equation}

For our next set of new inequalities, we will rely on Bochner's Theorem from
real analysis~\cite{bochner}. First, a complex-valued function $f$ on $\BR$ is called
{\it positive-definite} if for any choice of complex numbers $\alpha_1 , \alpha_2 ,
\cdots , \alpha_r$ and for any $x_1 , x_2 , \cdots , x_r \in \BR$, we have
\begin{equation}
\sum_{i,j=1}^r \alpha_i \overline{\alpha_j} f(x_i - x_j ) \geqslant 0 \; .
\end{equation}
Observe that consequences of positive-definiteness are: (i) $f(0) \geqslant 0$ (derived
from Eq.\,(30) with $r=1$), and (ii) $f(-x)=\overline{f(x)}$
and $f(0) \geqslant |f(x)| \;\; \forall x \in \BR$ (derived
from Eq.\,(30) with $r=2$). It turns out that positive-definiteness, along with the
normalization condition $\Phi (0)=1$, completely characterizes functions $\Phi$
that are expressible as the Fourier transform of a positive density of states $\rho (E)$
as in Eq.\,(18):

\noindent {\bf Bochner's Theorem}: A continuous complex-valued function $f:
\BR \longrightarrow \BC$ with $f(0)=1$ is the Fourier transform
$f(x)=\int e^{-i x E} \, \rho(E) \, dE$ of a finite, normalized, positive
Borel measure $\rho (E)$ if and only if $f$ is positive-definite.

\noindent Since Bochner's theorem completely characterizes a general
overlap function $\Phi (t)$, the (in general infinite) class
of inequalities
\begin{equation}
\sum_{i,j=1}^r \alpha_i \overline{\alpha_j} \, \Phi (t_i - t_j ) \geqslant 0 \;
\; \; \; \; \forall \; \alpha_1 , \alpha_2 ,
\cdots , \alpha_r \in \BC \, , \; \; t_1 , t_2 , \cdots , t_r \in \BR \; ,
\end{equation}
are the most general, universal set of inequalities constraining an overlap
function $\Phi (t)$ defined as in Eq.\,(2) for a pure state. Consequently,
every inequality  of the Margolus-Levitin-Mandelstam-Tamm-type constraining the
first zero $T_0$ of $\Phi (t)$ is necessarily
a consequence of Eqs.\,(31).

To derive some new examples of such inequalities for the first zero-crossing time
$T_0$ from Eqs.\,(31), observe that if $\Phi (t)$ is an
overlap function (hence is positive-definite), then Eqs.\,(18) and (25) combined with Bochner's
theorem imply that for any
positive integer $n$,
\begin{equation}
f(t) \equiv (-1)^n \frac{\Phi^{(2n)}}{\langle E^{2n} \rangle}
\end{equation}
is also a positive-definite
function satisfying $f(0)=1$. For ease of calculation, let us redefine
the hamiltonian as
\begin{equation}
H \longrightarrow H- \langle H \rangle \; ,
\end{equation}
so that the
average (first-moment) of $H$ vanishes: $\langle H \rangle =\langle E \rangle = 0$.
Taking $n=1$ in Eq.\,(32) and applying the basic consequence $1=f(0) \geqslant | f(t) |$
of Eqs.\,(30) gives
\begin{equation}
-\Phi'' (t) \leqslant \langle E^2 \rangle \; .
\end{equation}
Integrating Eq.\,(34) once from $t=0$ to $t$ and using $\Phi' (0) = -i \langle E \rangle
=0$ we obtain the inequality
\begin{equation}
-\Phi' (t) \leqslant \langle E^2 \rangle \, t \; .
\end{equation}
Integrating Eq.\,(35) once more from $t=0$ to $t=T_0$ (the first
zero-crossing) gives
\begin{equation}
1 \leqslant \langle E^2 \rangle \frac{{T_0}^2}{2} \; .
\end{equation}
After restoring to natural time units and undoing the rescaling Eq.\,(33), Eq.\,(36)
becomes the inequality
\begin{equation}
T_0 \geqslant \frac{\sqrt{2} \hbar}
{\sqrt{\langle \; ( E  - \langle E \rangle)^2 \, \rangle}} \; .
\end{equation}
\noindent Applying an entirely parallel stream of arguments and starting from $n=2,\; 3,\; \ldots$
in Eq.\,(32), we obtain the following infinite series of new inequalities
involving the first zero-crossing time $T_0$:
\begin{eqnarray}
\langle  \; ( E  - \langle E \rangle)^2 \,  \rangle
\frac{{T_0}^2}{2 \, \hbar^2 } & \geqslant & 1 \nonumber \\
\langle  \; ( E  - \langle E \rangle)^4 \,  \rangle \frac{{T_0}^4}{24 \, \hbar^4} & \geqslant & 
\langle  \; ( E  - \langle E \rangle)^2 \,  \rangle \frac{{T_0}^2}{2 \, \hbar^2 } - 1 \nonumber \\
\langle  \; ( E  - \langle E \rangle)^6 \,  \rangle \frac{{T_0}^6}{6! \, \hbar^6 } & \geqslant &
\langle  \; ( E  - \langle E \rangle)^4 \,  \rangle \frac{{T_0}^4}{24 \, \hbar^4 } -
\langle  \; ( E  - \langle E \rangle)^2 \,  \rangle \frac{{T_0}^2}{2 \, \hbar^2 } + 1 \nonumber \\
\vdots & & \vdots \nonumber \\
\langle  \; ( E  - \langle E \rangle)^{2n} \,  \rangle \frac{{T_0}^{2n}}{(2n)! \, \hbar^{2n} } & \geqslant &
\sum_{s=1}^{n} (-1)^s \langle  \; ( E  - \langle E \rangle)^{2(n-s)}
\,  \rangle \frac{{T_0}^{2(n-s)}}{[2(n-s)]! \, \hbar^{2(n-s)} }
\; .
\end{eqnarray}

~

{\noindent \bf Connection with thermodynamics:}
\renewcommand{\thefootnote}
{\ensuremath{\fnsymbol{footnote}}}
\addtocounter{footnote}{2}

As we argued above,
for a wide class of physical systems
the overlap function $\Phi (t)$ given by Eq.\,(18) is holomorphic on
the complex $t$-plane. It is straightforward to observe that at imaginary
times the overlap function is equal to the canonical partition function of
a thermodynamic system: for $\beta > 0$:
\begin{equation}
Z( \beta ) \equiv \Phi (-i \beta )
= \int e^{- \beta E} \, \rho(E) \, dE \; .
\end{equation}
Introducing the canonical probability distribution
\begin{equation}
\rho_c (E) \equiv \frac{e^{-\beta E} \, \rho (E)}{Z(\beta )}
\end{equation}
completes the correspondence with the thermodynamics of
a (in general, abstract) physical system whose density
of states is given by $\rho (E)$.\footnote{This system does not have
to be quantum mechanical; in general,
a classical system could be designed to have the density
of states $\rho (E) \, dE$ lying between its constant-energy shells
$\{ H= E \}$ and $\{H= E+dE \}$.}
For example, the thermodynamic entropy is given by
\begin{equation}
\frac{S}{k_B} \equiv
-\int\rho_c (E) \, \log \rho_c (E) \, dE
= \log Z(\beta ) + \beta \langle E \rangle_c - \langle \log \rho (E) \rangle_c
\; ,
\end{equation}
where $k_B$ is Boltzmann's constant, and
$\langle \cdots \rangle_c$ denotes expectation with respect to the canonical
probability distribution function $\rho_c (E)$:
$\langle \cdots \rangle_c \equiv \int \cdots \, \rho_c (E) \, dE$.
Real zeros of the overlap function $\Phi (t)$
correspond to pure-imaginary zeros of the partition function $Z( \beta )$.

The promise of the thermodynamic correspondence lies in the fact that the qualitative behavior of
thermodynamic systems have been extensively investigated, and contain a vast panoply of results and
techniques that might potentially translate to insights into the real-time behavior of
$\Phi (t)$ via the imaginary inverse-temperature correspondence with $Z( \beta )$.
For example, an impressive array of qualitative and quantitative results are
already known for the behavior of the complex zeros
of $Z( \beta )$ for a variety of lattice models in statistical
mechanics~\cite{biskup1, biskup2}. The condensation of
the complex zeros of the volume-scaled partition function onto the real axis
in the thermodynamic scaling limit is directly related to the phenomenon of phase transitions
according to the classical Lee-Yang theory~\cite{lee-yang}.



~~~~

\end{document}